\documentclass[twocolumn,a4paper,preprintnumbers,amsmath,amssymb,floatfix,showpacs]{revtex4}

\usepackage{graphicx}
\usepackage{dcolumn}
\usepackage{bm}

\begin{document}

\title{Vortex states in patterned exchange biased NiO/Ni samples}

\author{Pablo Asshoff}
\email[Electronic mail: ]{pablo.asshoff@ruhr-uni-bochum.de}
\author{Katharina Theis-Br{\"o}hl}
\author{Oleg Petracic}
\author{Hartmut Zabel}
\affiliation{Institut f{\"u}r
Experimentalphysik/Festk{\"o}rperphysik, Ruhr-Universit{\"a}t
Bochum, 44780 Bochum, Germany}

\begin{abstract}

We investigated the magnetization reversal of arrays of exchange
biased NiO/Ni squares with superconducting quantum interference
device magnetometry and micromagnetic simulations. The edges of the
squares were 0.5, 1.5, and 3.0~$\mu$m long. The NiO/Ni structures
exhibit vortexlike hysteresis loops and micromagnetic calculations
show that this feature is due to several vortices nucleating in the
islands. Furthermore, for the arrays with squares of 1.5~$\mu$m edge
length, the sign of the exchange bias field changes, as compared to
the same continuous NiO/Ni layer. We attribute the vortex nucleation
and the change of the exchange bias field to the interplay between
shape and unidirectional anisotropy.

\end{abstract}

\pacs{75.70.Kw, 75.30.Gw, 75.60.Jk, 75.75+a}


\maketitle

\section{INTRODUCTION}

One of the most renowned features of the exchange bias (EB) effect,
which results from the exchange coupling between ferromagnetic (F)
and antiferromagnetic (AF) layers, is a shift of the hysteresis loop
away from zero field. It is associated with a frozen-in
unidirectional anisotropy, induced, e.g. by cooling of the EB system
through the N\'eel temperature in an applied magnetic field. Due to
the intriguing properties underlying the EB phenomenon and its
relevance for magnetic recording applications it is under persistent
study since the seminal paper by Meiklejohn and Bean in 1956
\cite{MB1} (for recent reviews see Refs.~\cite{NoguesSchuller, Kiwi,
StampsReview, BerkowitzReview, RaduZabel}). Especially the
characteristics of EB in micro- to submicrometer dots are of
considerable interest with regard to the miniaturization of
industrially manufactured spin valves. Moreover, from a fundamental
viewpoint, it is of high importance to probe the effects of
confinement on the EB effect. Technical improvements in the
fabrication of nano- and micro\-structures in the past years have
made a closer study of EB in micro- to submicrometer dots
possible~\cite{NoguesNano}.

In exchange biased nanostructures in addition to the unidirectional
anisotropy and the intrinsic magnetocrystalline anisotropy of the AF
and F layers, the shape anisotropy is also significant. The
interplay between these anisotropies is crucial for domain formation
and remagnetization processes in nanostructures~\cite{Eisenmenger,
Temst2, PetracicDots}. While vortex formation sometimes is observed
in the F layer of exchange biased nanostructures~\cite{SortPRL}, the
unidirectional anisotropy may also lead to a collapse of the vortex
state that occurs in the same F nanostructure without being in
contact to an antiferromagnet~\cite{ClosureEBQuadrat}. An obvious
question is whether the opposite is also possible: vortex formation
in the EB system, but no vortex nucleation in the corresponding
unbiased F island.

In this paper, we investigate NiO/Ni EB systems. We analyze the
magnetic behavior of a continuous NiO/Ni bilayer film and of NiO/Ni
square islands of different sizes, both with (EB case) and without
(non-EB case) the existence of unidirectional anisotropy. For the
patterned samples, we see vortexlike hysteresis loops in the EB
case, whereas in the non-EB case several features of a vortex
hysteresis loop are missing. Micromagnetic simulations show that the
EB field can cause a vortexlike state that is not observed for the
same F microstructure.

\section{EXPERIMENTAL DETAILS}

The NiO/Ni samples were fabricated by UHV ion beam sputter
deposition and subsequent plasma oxidation. After heating the
MgO(001) substrate to 800~\ensuremath{^\circ}C for cleaning its
surface, Ni was deposited at a substrate temperature of
100~\ensuremath{^\circ}C with a sputter rate of 0.03 nm/s. The
sample was then annealed at 300~\ensuremath{^\circ}C for several
hours promoting Ni(001) growth. For a similar sample, Lukaszew et
al.~\cite{Interdiffusion} report on an interdiffusion process of the
oxygen from the MgO substrate to the Ni, thus creating NiO. While
they found evidence of a NiO/Ni EB system, our superconducting
quantum interference device (SQUID) measurements of the as-prepared
samples never showed an EB effect.

To generate the NiO layer, we used rf plasma oxidation. Test
experiments with other oxidation methods showed that plasma
oxidation appears to be the only method to obtain exchange biased
films by means of oxidizing a Ni layer, probably because a minimum
film thickness of the AF layer is required for nonzero EB fields.
The plasma oxidation was performed at 200~\ensuremath{^\circ}C with
an rf power of 15~W in a separate oxidation chamber, filled with
pure oxygen gas at a pressure of $6 \times 10^{-2}$ mbar.

\begin{figure} \centering
\includegraphics[clip=true,keepaspectratio=true,width=0.8\linewidth]{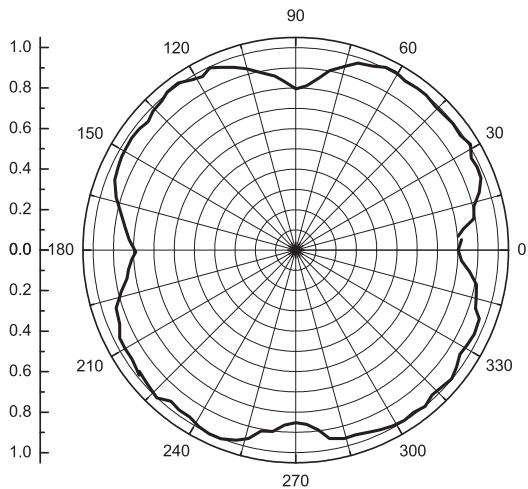}
\caption{\label{Ni_NiO_MOKE}Remanent magnetization normalized to the
saturation magnetization as a function of the sample rotation angle,
measured with MOKE. A four-fold magnetocrystalline anisotropy is
observed.}
\end{figure}

\begin{figure} \centering
\includegraphics[clip=true,keepaspectratio=true,width=0.8\linewidth]{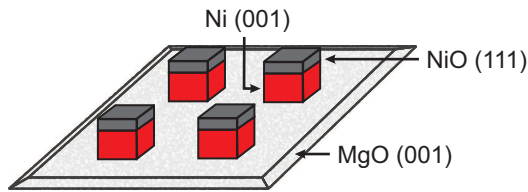}
\caption{\label{SchemeDots}Schematic drawing of the EB patterns (not
to scale).}
\end{figure}

As explored by small angle x-ray reflectivity and wide angle x-ray
diffraction, the resulting Ni layer has a thickness of 4.8~nm and
shows a (001) out-of-plane texture. The NiO thickness amounts to
2.0~nm with a (111) out-of-plane texture.

We determined the magnetocrystalline anisotropy of the Ni film with
a standard longitudinal magneto-optical Kerr effect (MOKE) setup
\cite{MOKE} and observed a four-fold anisotropy, as shown in Fig.
\ref{Ni_NiO_MOKE}. The figure displays the value of the remanence
normalized to the saturation value as a function of the relative
sample rotation angle. The sample was rotated in steps of
5\ensuremath{^\circ}. The easy axis lies parallel to the in-plane
[1~1~0] direction, in agreement with reports in the
literature~\cite{Hwang}.

On the same bilayer, square microstructures with different diameters
were defined by electron beam lithography. The edges of the squares
were aligned along the magnetic easy axis of the continuous Ni film,
,i.e., the [1~1~0] direction. Electron beam lithography was
performed with negative photoresist and a FEI Quanta 200 field
emission gun (FEG) scanning electron microscope controlled by the
Raith ELPHY QUANTUM 4.0 software. After the lithography step, the
excess F and AF layers were removed by ion beam etching. The
remaining squares had edge lengths of 0.5~$\mu$m, 1.5~$\mu$m, and
3.0~$\mu$m.

\begin{figure} \centering
\includegraphics[clip=true,keepaspectratio=true,width=0.8\linewidth]{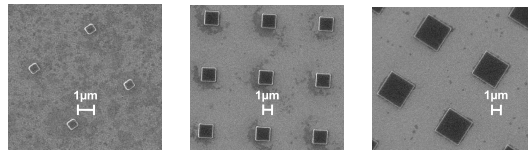}
\caption{\label{SEMimages}SEM images of the arrays of square dots
with edge lengths of 0.5, 1.5, and 3.0~$\mu$m, respectively, and a
constant spacing of 4~$\mu$m between the dots.}
\end{figure}

A sketch of the laterally patterned films is shown in
Fig.~\ref{SchemeDots}. While the diagram displays just four dots, on
the sample approximately $10^5$ squares were arranged in order to
have a magnetic signal detectable by a SQUID magnetometer.
Representative scanning electron microscope (SEM) images of the
patterns are shown in Fig.~\ref{SEMimages}.

Subsequently, SQUID magnetometry was used to characterize the EB
properties of the continuous bilayer and the patterned films. We
employed a rf-SQUID magnetometer by Quantum Design in the
reciprocating sample option mode, which is more suitable for weak
magnetic signals than the dc stepped-scan technique.

\section{EXPERIMENTAL RESULTS}

For NiO films with a thickness of a few nanometers the N\'eel
temperature is expected to be considerably reduced as compared to
the bulk value of 523~K \cite{NiONeelTemperature}. Alders et al.
report a decrease to 295~K for 5 monolayers of NiO(001) (2.1~nm)
\cite{NiOThinFilms1}. For our NiO/Ni samples we proved that the
blocking temperature is below 400~K, which makes them suitable for
performing the field-cooling procedure in the SQUID magnetometer
that provides a maximum sample temperature of 400~K. The following
procedure was employed: We measured the hysteresis loop at 10~K
after field cooling from 400~K in a field of 0.3 T. Then we repeated
this measurement after rotating the sample by 90\ensuremath{^\circ}
degrees. If there were frozen-in spins from the previous
field-cooling procedure (at the starting angle), then the hysteresis
loops are expected to differ, indicating that the blocking
temperature was above 400~K. However, in the experiment, we observed
identical hysteresis loops at different angles. Therefore, we
conclude that in our NiO/Ni samples the blocking temperature is
below 400~K.

\begin{figure} \centering
\includegraphics[clip=true,keepaspectratio=true,width=.67\linewidth]{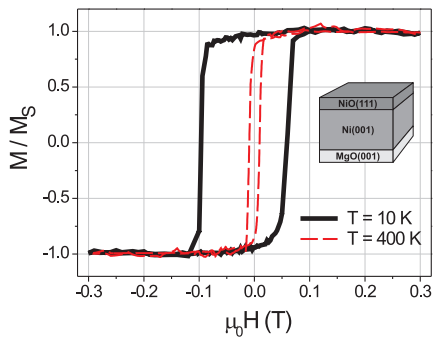}
\caption{\label{NiONi_unstr}SQUID hysteresis loop of the continuous
NiO/Ni film measured at 10~K after field cooling from 400~K in a
field of 0.3~T (solid curve) and the hysteresis loop in the non-EB
case at 400~K (dashed curve).}
\end{figure}

\begin{figure} \centering
\includegraphics[clip=true,keepaspectratio=true,width=0.67\linewidth]{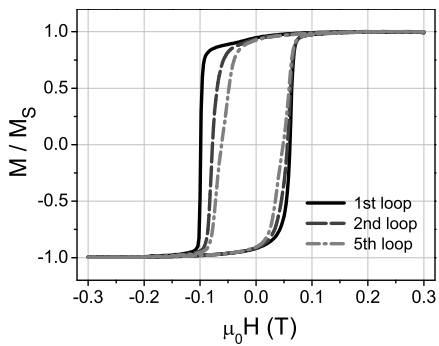}
\caption{\label{TrainingEffect}SQUID hysteresis loops of the
continuous NiO/Ni film measured at 10~K after field cooling from
400~K in a field of 0.3~T for the first, second and fifth hysteresis
loops.}
\end{figure}

Accordingly, the hysteresis loop of the continuous NiO/Ni film was
measured at 10~K after field cooling from 400~K in a field of 0.3~T.
The field was applied along the easy axis and its value exceeded the
saturation magnetization. In Fig.~\ref{NiONi_unstr} the resulting
data are displayed, together with the hysteresis loop for the non-EB
case, measured at 400~K, i.e., above the blocking temperature. After
field cooling, the continuous NiO/Ni sample exhibits an EB field of
$H_{EB}=-18.75$~mT, and the hysteresis loop reveals a slight
asymmetry in the ascending and descending branches: The lower right
part of the hysteresis is more rounded than the upper left part. Due
to this observation, we analyzed the training effect of the EB
system. In Fig. \ref{TrainingEffect} the virgin hysteresis loop and
the magnetization reversal for the second and the fifth cycle are
shown, again measured at 10 K after field cooling starting at 400 K.
Obviously, only the virgin curve is asymmetric, whereas the trained
hysteresis curves are centered around a point on the field axis set
by the EB field.

\begin{figure} \centering
\includegraphics[clip=true,keepaspectratio=true,width=.7\linewidth]{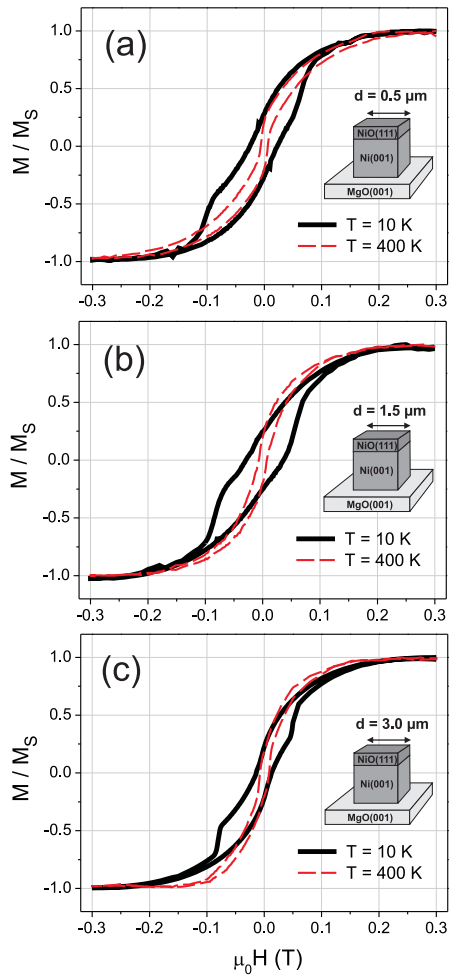}
\caption{\label{NiONi_patterned}SQUID hysteresis loops of the
patterned NiO/Ni samples: square dot arrays with 0.5~$\mu$m edge
lengths (a), 1.5~$\mu$m edge lengths (b), and 3.0~$\mu$m edge
lengths (c). The solid curves correspond to measurements in the EB
case at 10~K, and the dashed curves to the hysteresis loops of the
system at 400~K, i.e., above the blocking temperature of the thin
NiO film.}
\end{figure}

The hystereses for the patterned samples are shown in
Fig.~\ref{NiONi_patterned}, where the field was applied along the
edges of the squares, i.e., along the easy axis according to the
continuous film. The shape of the hysteresis loop at 400~K changed
remarkably compared to the continuous NiO/Ni film and clearly
indicates a domain state at remanence. Moreover, the magnetic
hysteresis loops at 10K (solid lines) show considerable divergence
of the shape, coercivity and EB field magnitude from the
corresponding data for the continuous NiO/Ni film. It is also
noticeable that in the EB case all hysteresis loops of the patterned
samples show kinks in each branch before reaching saturation and
exhibit an asymmetric shape. However, the shape of the hysteresis
loops measured in the non-EB case is symmetric and no kinks are
visible. Altogether, the hysteresis loops indicate that in the EB
case the square microstructures have a domain configuration similar
to a vortex state. In the non-EB case a different remagnetization
process is observed, although it should be noted that the loop is
slightly more narrow close to remanence than for higher fields.

Another remarkable feature is the behavior of the EB fields given in
table~\ref{TableNiONiEB}. For the squares with 0.5 and 3.0~$\mu$m
edge lengths, the EB field is reduced to -1.25 and -0.75~mT,
respectively. Surprisingly, for the squares with 1.5~$\mu$m edge
length, the hysteresis is shifted to a positive EB with
$H_{EB}=+3.3$~mT. This positive EB was observed for several samples
with squares of 1.5~$\mu$m edge size, so that we can exclude
individual structural and lithographical properties of the sample.
The positive EB appears to be related to changes in the
remagnetization behavior and it is thus not a fundamentally new
feature of EB in microstructures, as described below.

\begin{table}
\begin{center}
\begin{tabular}{|c|c|c|c|}
\hline Edge size of squares& $H_{C_{1}}$& $H_{C_{2}}$
 & $H_{EB}$ \tabularnewline
($\mu$m)& (mT)&  (mT)& (mT)
  \tabularnewline \hline \hline Continuous
film& -96.5& +59.0& -18.75\tabularnewline \hline 3.0 & -14.0& +12.5&
-0.75\tabularnewline \hline 1.5 & -29.4& +36.0& +3.3\tabularnewline
\hline 0.5& -25.0& +22.5& -1.25\tabularnewline \hline
\end{tabular}
\caption{Coercive fields and EB fields for the continuous NiO/Ni
bilayer and the NiO/Ni microstructures calculated from the classical
EB formula.} \label{TableNiONiEB}
\end{center}
\end{table}

\begin{table}
\begin{center}
\begin{tabular}{|c|c|c|c|}
\hline Edge size of squares& $H_{C_{1}}$& $H_{C_{2}}$
 & $H_{EB}$ \tabularnewline
($\mu$m)& (mT)&  (mT)& (mT)
  \tabularnewline \hline \hline Continuous film& -96.5
($H_{C_1}$)& +59.0 ($H_{C_2}$)& -18.75\tabularnewline \hline
 3.0 & -76.3& +47.1&
-14.6\tabularnewline \hline 1.5 & -70.7& +44.3& -13.2
\tabularnewline \hline 0.5 & -90.2& +55.4& -17.4\tabularnewline
\hline
\end{tabular}
\caption{Position of the kinks for the microstructures and EB fields
resulting from averaging these values (only for the continuous
NiO/Ni bilayer the EB field listed is calculated from the classical
EB formula).} \label{TableKinks}
\end{center}
\end{table}

\section{MICROMAGNETIC SIMULATION}

To determine the origin of the vortexlike shaped hysteresis loops,
micromagnetic calculations with the OOMMF simulation software
package were performed~\cite{Oommf}. We modeled a square dot with an
edge length of 0.5~$\mu$m consisting of nickel (saturation
magnetization $M_S=0.49$~MA/m and exchange constant $A=3.4$~pJ/m). A
cubic magnetocrystalline anisotropy~\cite{footnote1} with the easy
axes along [1~1~0], [-1~1~0] and [0~0~1] and an anisotropy constant
of $-5.7$~kJ/m$^{3}$ were assumed. The EB field was simulated with
locally pinned AF spins distributed randomly across 10\% of the
area~\cite{PetracicBidomain}.

\begin{figure} \centering
\includegraphics[clip=true,keepaspectratio=true,width=0.93\linewidth]{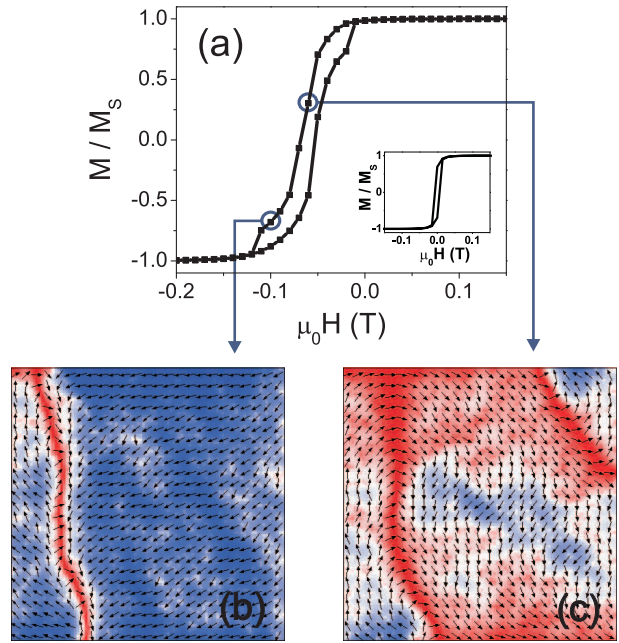}
\caption{\label{simulation}Results of micromagnetic simulations for
a square dot of 0.5~$\mu$m edge length. (a) Hysteresis loop for the
EB case and two characteristic domain configurations [(b) and (c)]
related to the field values indicated in the hysteresis. The inset
in (a) shows the hysteresis loop of the square dot in the non-EB
case.}
\end{figure}

The hysteresis loop obtained from the micromagnetic model when the
field is applied along the edge of the structure is shown in
Fig.~\ref{simulation}(a). The inset shows the simulation result for
the non-EB case, where no pinned AF spins are assumed. It is
symmetric and no vortex state occurs. The more rounded experimental
hysteresis loop is most likely due to the sequential switching of
individual dots with a distribution of coercive fields. In the EB
case, the hysteresis loop has a characteristic shape resembling the
experimental data from Fig.~\ref{NiONi_patterned}. In particular,
the narrowing close to the coercive fields and the kinks in the
ascending and descending branches of the hysteresis loops are
properly reproduced. However, compared to the experimental data, the
loop shows a pronounced EB field $H_{EB}$, and the remagnetization
process occurs within a rather small field range. As previously
shown, we attribute this deviation to the spin-glass behavior of the
interface between the F and the AF layer. It implies that some of
the spins at the interface are frozen in and contribute to the EB
shift, while others reverse concurrently with the F layer, as soft
x-ray resonant magnetic scattering (XRMS) experiments
documented~\cite{RaduExpProofSpinGlass}. If the latter behavior of
the interface spins is taken into account, the hysteresis loop of
the EB system broadens and the EB field is reduced, as shown by
numerical calculations~\cite{RaduZabel}. For the micromagnetic model
employed in this paper, these features of the interface spins are
not considered, which describes the deviation between the simulated
and experimental magnetization reversals qualitatively.

For two characteristic points of the hysteresis loop the domain
configuration is displayed in Fig.~\ref{simulation}. For the more
narrow parts of the hysteresis loop close to remanence domain
configurations as in Fig.~\ref{simulation}(c) occur, which contain
vortexlike states. The kinks emerge due to the collapse of the
vortexlike states and the formation of a 360\ensuremath{^\circ}
domain wall across the square structure [Fig.~\ref{simulation}(b)].

In the simulation we could also see that vortexlike hysteresis loops
only arise if a certain balance between the values of the EB field
and the shape anisotropy energy exists. By varying either the
unidirectional anisotropy strength or the structure size in the set
of simulation parameters, we obtained hysteresis loops, which
resemble all of those displayed in Fig.~\ref{NiONi_patterned}, apart
from an overall shift along the field axis. We observed hysteresis
loops with kinks located symmetrical to the hysteresis's center, but
at different field values. Moreover, the positions of the kinks can
move such that the upper and the lower kink are encountered at
fields asymmetrical to the center of the hysteresis. This might
explain the appearance of a positive EB field, being due to a kink
forming in the ascending hysteresis branch at the coercive field,
but at a different field in the descending branch. However, in the
simulation, these features show only a qualitative agreement with
the hysteresis loops shown in Fig.~\ref{NiONi_patterned}. In
particular, the varying EB field leads to a nonconstant shift of the
hysteresis loop. As before, we attribute these deviations to spin
disorder at the NiO/Ni interface.

\section{DISCUSSION}

\subsection{Continuous NiO/Ni film}

For the continuous NiO/Ni film, the shape of the first hysteresis
loop (virgin state) is typical of a reversal through nucleation of
domain walls and domain wall motion. The second and subsequent
reversals indicate that a reversal through rotation of magnetization
occurs. For CoO/Co EB systems, the asymmetric shape of the first
hysteresis loop has been observed before in, e.g.,
Refs.~\cite{FlorinPRB} and~\cite{McCordAsymmetricReversal}, where
the formation of magnetic domains at the interface subsequent to the
first magnetization reversal was reported. In theoretical studies,
this training effect was related to irreversible changes taking
place during the first reversal at the F/AF interface and in the AF
layer~\cite{HoffmannTraining,BinekTraining,RaduZabel}. A more
thorough review of the training effect within EB models is given in
Ref.~\cite{RaduZabel}, including the spin-glass model of EB, which
is able to explain many of the observed experimental features
associated with the training effect of EB in continuous bilayers.

\subsection{Square microstructures}

The hysteresis loops in the EB case differ considerably from those
in the non-EB case. After the field-cooling procedure, the
hysteresis loops exhibit kinks before reaching saturation and the EB
field is reduced. This tightening of the hysteresis at coercivity is
a feature reminiscent of a vortex state~\cite{BelegFerro1}. However,
in contrast to vortex states in circular islands, the squares appear
not to have a well defined nucleation field.

Modified magnetization reversal processes for EB patterns, as
compared to continuous exchange biased films, were previously
reported. E.g., Eisenmenger et al.~\cite{Eisenmenger} reported on
strongly asymmetrically shaped hysteresis loops of exchange biased
Fe/FeF$_2$ circular nanostructures. In a study by Li et
al.~\cite{PetracicDots} a vortex state within exchange biased
Fe/FeF$_2$ nanostructures was identified by combined SQUID
investigations and micromagnetic simulations. The hysteresis loops
above the N\'eel temperature resemble those in the EB case strongly.
Kato et al.~\cite{ClosureEBQuadrat} imaged square MnIr(20
nm)/NiFe(20 nm) and NiFe(20 nm) nanostructures with MFM that exhibit
closure domains. But closure domains for the MnIr/NiFe structures
were only found when the corresponding NiFe element also showed a
closure domain. This is consistent with the intuitive argument that
an additional contribution from unidirectional anisotropy should
destroy the highly symmetrical vortex state.

However, this argument does not hold for the patterned samples
studied here. Both in the micromagnetic simulation and in the
experimental data the vortex states are only present in the EB case
but not in the non-EB case. Thus, the change in the magnetization
reversal including the kinks in the hysteresis loops must be due to
the additional energy contribution from unidirectional anisotropy.
As can be seen from the micromagnetic simulation, the local EB acts
as a nucleation center for several vortices within the squares with
0.5~$\mu$m edge length. A similar behavior is expected for the other
patterned samples. Moreover, from the numerical calculations, it
becomes evident that the balance between shape and unidirectional
anisotropy is crucial for the position of the kinks. This is
consistent with the observation for the square islands with
1.5~$\mu$m edge length, where the kinks are not symmetrical to the
center of the hysteresis. Additionally to the effects on the shape,
the EB field pushes the whole vortexlike hysteresis loop to the
negative side of the field axis, acting similarly as in a continuous
film. Thereby, the kinks are not encountered at positions equally
distant to zero field any more.

Related to the remagnetization behavior is the positive EB field
found for the NiO/Ni islands with 1.5 $\mu$m edge length. A positive
EB has been seen before in continuous films and is usually
attributed to uncompensated frozen-in spins in the AF layer, which
align parallel to the field direction in high enough external
fields. For EB islands, a positive shift should rather be related to
the shape change in the hysteresis. With a variation of the size of
the patterns, the shape anisotropy also changes. As a result of the
interplay between shape and unidirectional anisotropy, the
hysteresis loop is then different in shape. Especially if the kinks
are encountered at positions not symmetrical to the center of the
hysteresis, as is the case for the islands with 1.5 $\mu$m edge
length, the coercive fields are affected. Considering that on the
length scale discussed the unidirectional anisotropy strength should
remain constant, as discussed below, the shape anisotropy should
therefore cause the positive EB.

Consequently, with the magnetization reversal being determined by
shape anisotropy, it is misleading to use the classical formula for
the EB field and relate it to the unidirectional anisotropy energy
in the system~\cite{ModEBfield}. In the classical EB formula
$H_{EB}= \left(H_{C1}+H_{C2}\right)/2$ only the coercive fields of
the ascending and descending branches $H_{C1}$ and $H_{C2}$ are
considered. We propose to use the field position of the kinks to
describe possible shifts of the hysteresis. This appears to be a
more appropriate procedure to determine the unidirectional
anisotropy energy, since the kinks of a typical vortex hysteresis
loop have equal distance to the center of the hysteresis. Although
multivortex states develop in the samples, which leads to deviations
from hysteresis loops typical of a single vortex state, this
procedure results in an almost constant EB field, as shown in
table~\ref{TableKinks}. The values for the EB field in the table
were determined by averaging the field axis values of the left and
right kinks. Furthermore, the value of the EB field for the
continuous film determined by the classical EB formula is comparable
to the EB fields for the patterned samples as obtained by averaging
over the positions of the kinks. This indicates that the
unidirectional anisotropy energy is constant for the different
samples, which is in agreement with the fact that the AF domain size
in the system studied should be rather small~\cite{Radu2008,Fraune}.
Thus, the local EB at the F/AF interface is not affected by the
patterning process down to 0.5 $\mu$m and the unidirectional
anisotropy energy remains constant for all the samples. However, the
combination of local EB and shape anisotropy has a dramatic effect
on the shape of the hysteresis in NiO/Ni microstructures.

\section{SUMMARY AND CONCLUSION}

We investigated the EB effect in NiO/Ni micro- and
submicrostructures. Due to the interplay between shape and
unidirectional anisotropy several interesting effects arise. First,
vortex formation is more distinctive in the exchange biased square
islands than in the corresponding unbiased F islands. Second, the
shape of the whole hysteresis is affected by a combination of
unidirectional anisotropy and shape anisotropy. Third, as a
consequence, the classical formula for the EB field is not
appropriate for the description of EB in structures containing a
vortex state. Misleading positive EB fields may result from this. A
modified procedure for evaluating the EB field is proposed, which
takes the change of the shape into account.

\section*{ACKNOWLEDGEMENT}

We would like to thank T. Eim\"uller, F. Radu, A. Remhof, P. Szary,
and K. Westerholt for helpful discussions. This work was supported
by SFB 491 of the Deutsche Forschungsgemeinschaft: "Magnetic
Heterostructures: Spinstructure and Spintransport", which is
gratefully acknowledged.

\bibliographystyle{apsrev}

\begin{thebibliography}{27}
\bibitem{footnote1}
Note that the result remains qualitatively similar if [1~1~1],
[1~1~-1] and [1~-1~-1] are defined as easy axes, corresponding to
the configuration in bulk Ni.
\end{thebibliography}

\end{document}